\documentclass[12pt, onecolumn]{IEEEtran}
\usepackage{amsmath}
\usepackage{amssymb}
\usepackage{algorithm}
\usepackage{algorithmic}
\usepackage[tight]{subfigure}
\usepackage{cite}
\usepackage{makecell}
\usepackage{multirow}
\usepackage{url}
\usepackage{breakurl}
\usepackage[breaklinks]{hyperref}
\usepackage{graphicx}
\usepackage{dblfloatfix}
\usepackage{booktabs}

\begin{document}

\title{Edge Network-Assisted Real-Time Object Detection Framework for Autonomous Driving}

\author{Seung-Wook Kim, Keunsoo Ko, Haneul Ko,~\IEEEmembership{Member,~IEEE}, and Victor C. M. Leung,~\IEEEmembership{Fellow,~IEEE}

\thanks{Seung-Wook Kim, Keunsoo Ko, and Haneul Ko are with Korea University (corresponding author: Haneul Ko); Victor C. M. Leung is with Shenzhen University and the University of British Columbia.}

\thanks{This paper is accepted in IEEE Network.}
}

\maketitle

\begin{abstract}
Computer vision tasks such as object detection are crucial for the operations of autonomous vehicles (AVs). Results of many tasks, even those requiring high computational power, can be obtained within a short delay by offloading them to edge clouds. However, although edge clouds are exploited, real-time object detection cannot always be guaranteed due to dynamic channel quality. To mitigate this problem, we propose an edge network-assisted real-time object detection framework~(EODF). In an EODF, AVs extract the region of interests~(RoIs) of the captured image when the channel quality is not sufficiently good for supporting real-time object detection. Then, AVs compress the image data on the basis of the RoIs and transmit the compressed one to the edge cloud. In so doing, real-time object detection can be achieved owing to the reduced transmission latency. To verify the feasibility of our framework, we evaluate the probability that the results of object detection are not received within the inter-frame duration (i.e., outage probability) and their accuracy. From the evaluation, we demonstrate that the proposed EODF provides the results to AVs in real-time and achieves satisfactory accuracy.
\end{abstract}

%\begin{keywords}
%Autonomous driving, autonomous vehicle, object detection, edge cloud, cloud computing.
%\end{keywords}

\IEEEpeerreviewmaketitle
\newpage

\section{Introduction}

Autonomous vehicles~(AVs) have paved the way in the new era of smart mobility based on a combination of various types of technologies such as cameras, radars, lidars, and computer vision~\cite{Yaqoob20}. AVs mitigate the users' driving burden by executing intelligent operations. They can reduce traffic congestion by managing the traffic flow, which leads to low energy consumption and environment preservation. In addition, AVs assist elderly and disabled people by providing reliable and safe transportation systems. Owing to these tremendous benefits, there is an increasing interest in AVs in industrial and academic fields.

Since the tasks for AVs including object detection require high computational power, expensive computational units~(e.g., GPU) should be built into vehicles to complete these tasks within a short duration.\footnote{For safe and exquisite operations of AVs, the short task completion time must be guaranteed.} Meanwhile, to fully achieve the advantages of AVs, most of the vehicles should be automated. For example, for congestion control, optimal traffic distribution can be achieved only when all vehicles on the roads are autonomous and under the control of a central entity. However, several potential customers hesitate in purchasing AVs due to their high price, which implies that it is impossible to transit from the current transportation system to a fully autonomous transportation system within a short period of time. To reduce the price and boost the supply of AVs, many works~(such as edge computing, stereo computer vision, and cooperative driving) have been investigated in the literature~\cite{Liu19,Chen19}. One of the most promising solutions is exploiting the edge clouds~\cite{Liu19}. Specifically, if vehicles offload their tasks that require high computational power to edge clouds, they can obtain the task results without any expensive computational units. For example, a vehicle can transmit an image to an edge cloud for object detection. After receiving the image, the edge cloud can detect objects using its sufficient computing power and return the result to the vehicle. For the edge cloud solution, a low transmission latency must be guaranteed, which depends on the channel quality.\footnote{Even though edge clouds are exploited, if the channel quality is poor and/or the size of the offloaded image is large, the low transmission latency cannot always be guaranteed.} However, the previous works~\cite{Guo20,Gao20} to perform object detection on the edge cloud do not consider the channel quality when offloading the image.

%Figure~\ref{fig:Motivation_example} illustrates a motive example for this article. Before transmitting the image to the edge cloud, a base station~(BS) and AV check the channel quality by exchanging control messages about the channel status information~(CSI) to select the appropriate modulation method and code rate~(steps 1-2). By exploiting the selected modulation method and code rate, the AV transmits the image to the edge cloud through the BS~(step 3). However, when the channel quality is very poor and/or the size of the offloaded image is large, the image arrives at the edge cloud with long latency. The edge cloud can perform OD and transmit its result only after this long latency~(steps 4-5). This causes a long operational latency, denoted by $T$, that may be larger than the inter-frame duration~(IFD), i.e., 1/30 sec. To mitigate this problem, a sophisticated offloading framework should be developed.

%\begin{figure}
%\includegraphics[width=12cm]{Figure/Operational_example2.eps} \centering
%\caption{Motive example.}
%\label{fig:Motivation_example}
%\end{figure}

In this article, we developed an edge network-assisted real-time object detection framework~(EODF) for autonomous driving. In the proposed framework, an AV verifies whether the channel quality is sufficient to transmit the original image to the edge cloud and receive the detection results within the inter-frame duration. If the channel quality is not sufficient, the AV extracts regions of interests~(RoIs) using a low complexity algorithm, that is, the spectral residual visual saliency~(SRVS) model~\cite{HouCVPR2007}.\footnote{Note that RoI extraction is different from object detection. RoI extraction just provides the segments or the bounding box without the identification of an object that can be determined by object detection.} Intuitively, since most of the important contextual information on objects would be included in the RoIs, the AV can compress the original image on the basis of the extracted RoIs. Specifically, it simply retains the image data in the RoIs and discards the rests. Then, the compressed image is transmitted to the edge cloud. This can help in significantly reducing the transmission latency, and therefore real-time object detection can be achieved even when the channel quality is poor. To verify the feasibility of the proposed EODF, we evaluate the probability that the results are not received within the inter-frame duration (i.e., outage probability) and the accuracy of the object detection results (i.e., average precision). From the evaluation, we demonstrate that the EODF can provide over $99\%$ results to the AV at 30 fps due to the reduced transmission latency. In addition, it can be shown that the EODF can achieve over a mean average precision of $84\%$ on the public dataset for autonomous driving because the EODF compresses the image without the significant loss of the important contextual information of the image.

The contribution of this article can be summarized as follows:  1) The EODF that adaptively compresses images according to the channel condition to ensure the accurate object detection in real-time can reduce the price of AVs and boost their supply, indicating that their social and economic benefits can be fully achieved in the near future; 2) Since the feasibility of the EODF is sufficiently verified in terms of the outage probability and detection accuracy, it can be used in practical systems; and 3) Extensive evaluation results based on real measurement values are presented and analyzed under various environments, which provide valuable guidelines for designing edge cloud-assisted autonomous driving systems.

The remainder of this article is organized as follows. In the next section, background knowledge is summarized. Then, the proposed EODF is elaborated, and the evaluation results are presented. Finally, we discuss future research directions, followed by the concluding remarks.

\section{Background}
\label{Sec:Background}

\subsection{Salient Region Detection}

RoIs are the regions that contain different levels of saliency in terms of color, texture, and shape within an image. Such regions include the important contextual information of objects in the image. Salient region detection is one of the representative methods to extract the RoIs. In salient region detection, an agent generates a saliency map by scoring pixels based on the possibility of whether objects exist or not. That is, in the saliency map, pixels, where objects are likely (not) to exist in the image, have high (low) values. Therefore, the agent can easily extract the RoIs based on the scores in the saliency map. Various studies have been conducted to generate confident saliency maps and accomplished great achievements~\cite{HouCVPR2007,borji2019salient}.

\begin{figure}[]
\centering
	{
	\subfigcapskip = -12pt
		\subfigure[]{
		\includegraphics[width =0.4\textwidth]{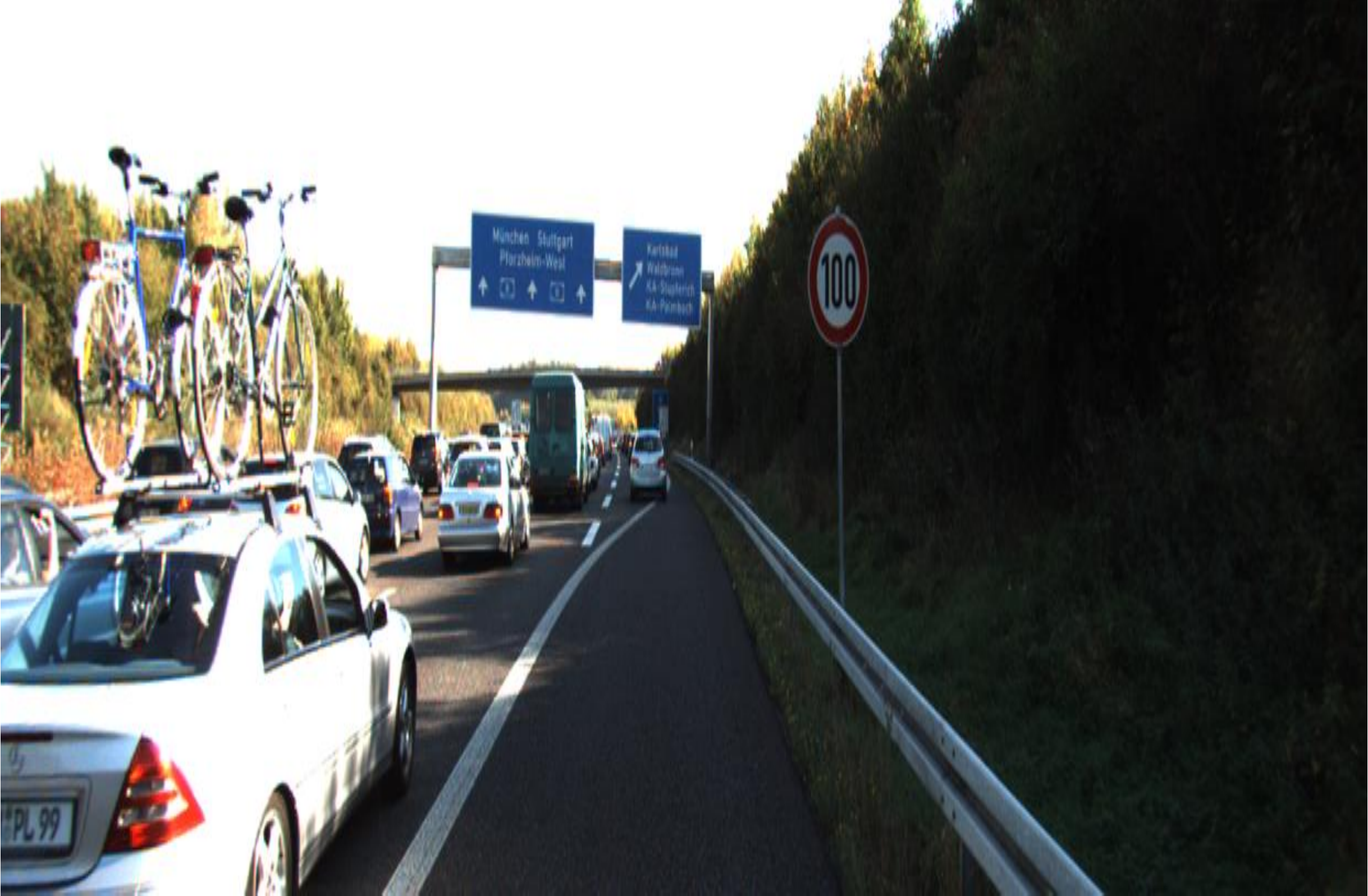}}
		\subfigure[]{
		\includegraphics[width=0.4\textwidth]{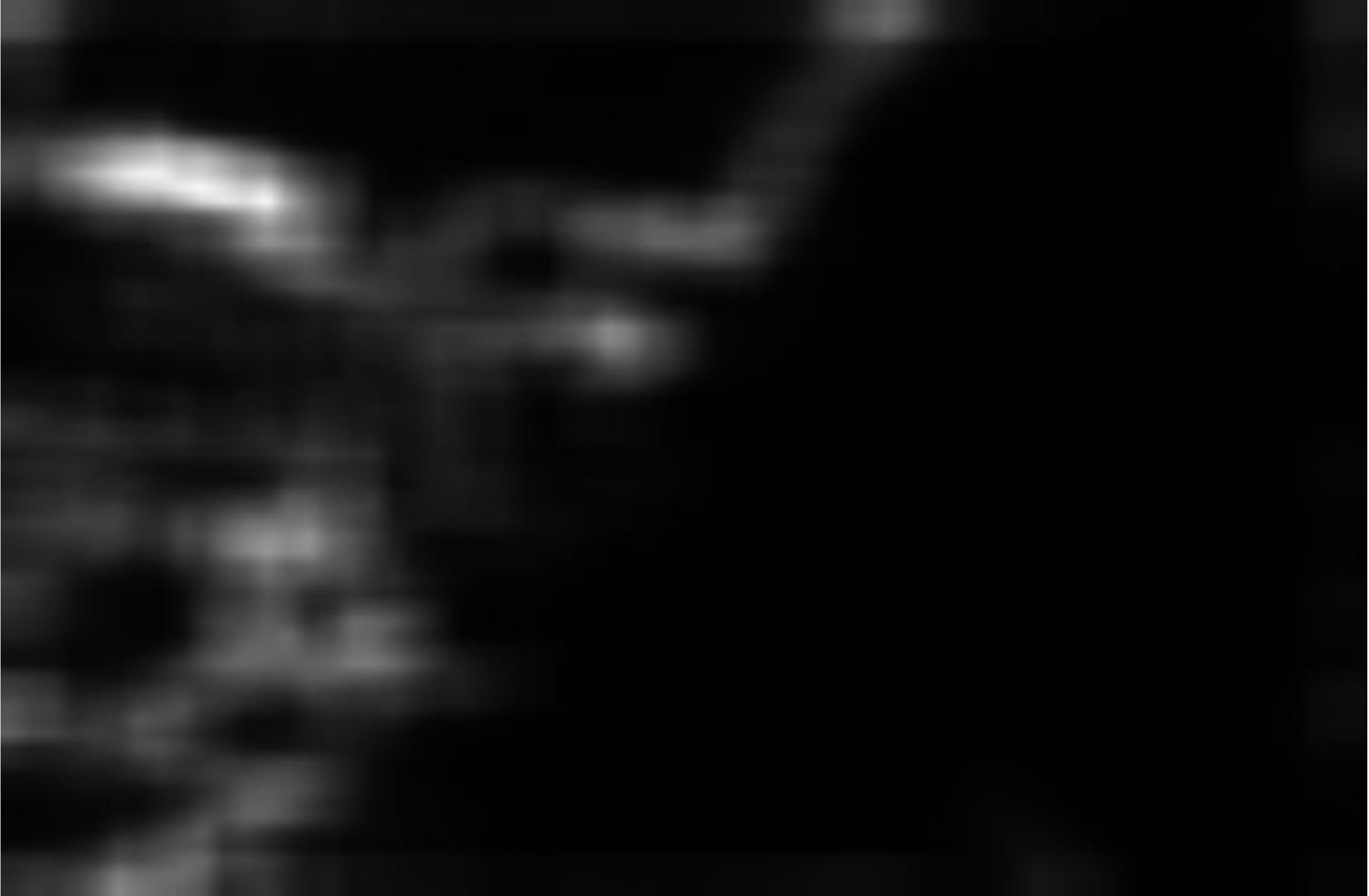}}\\
		\subfigure[]{
		\includegraphics[width =0.4\textwidth]{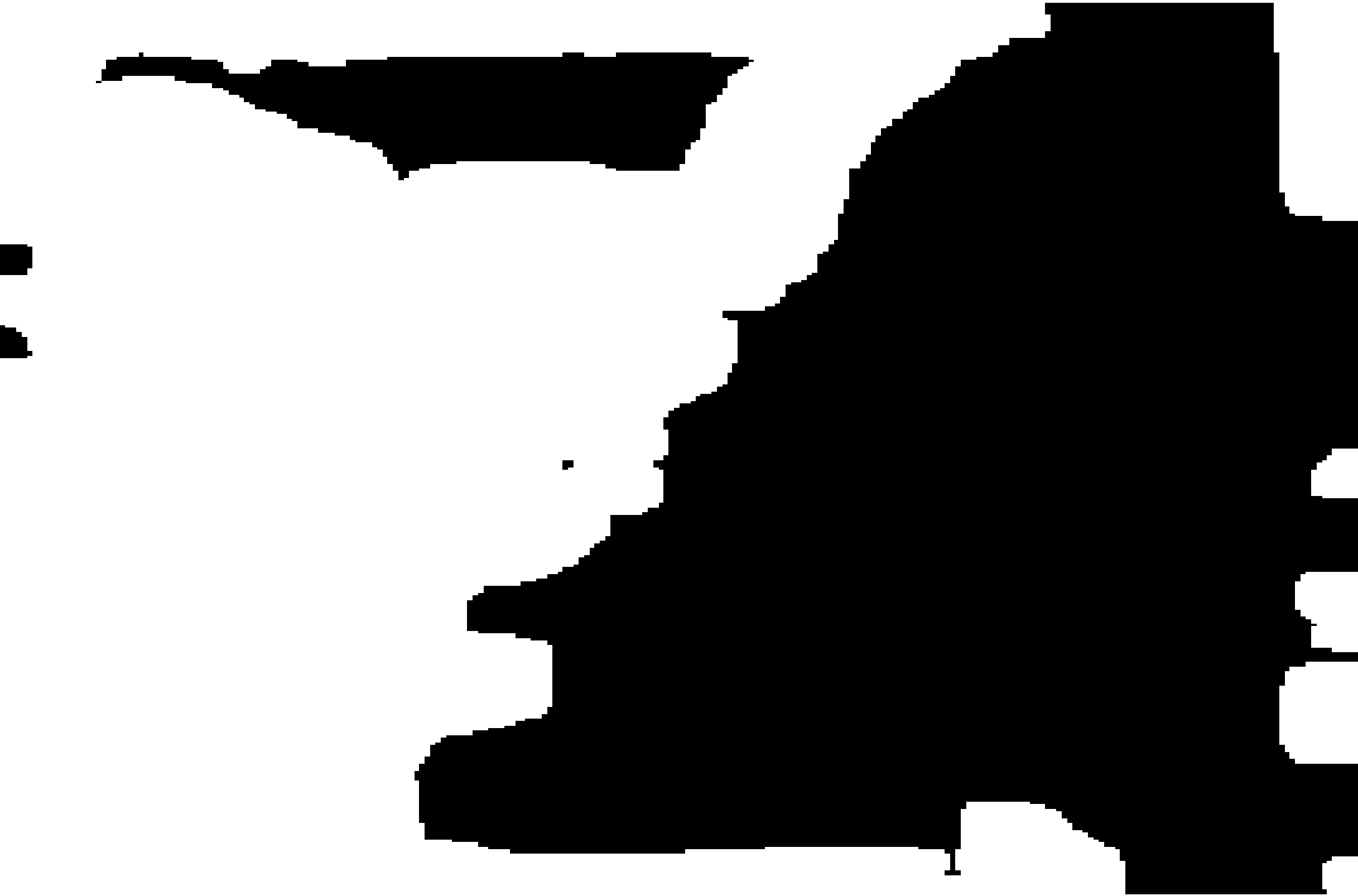}}
		\subfigure[]{
		\includegraphics[width=0.4\textwidth]{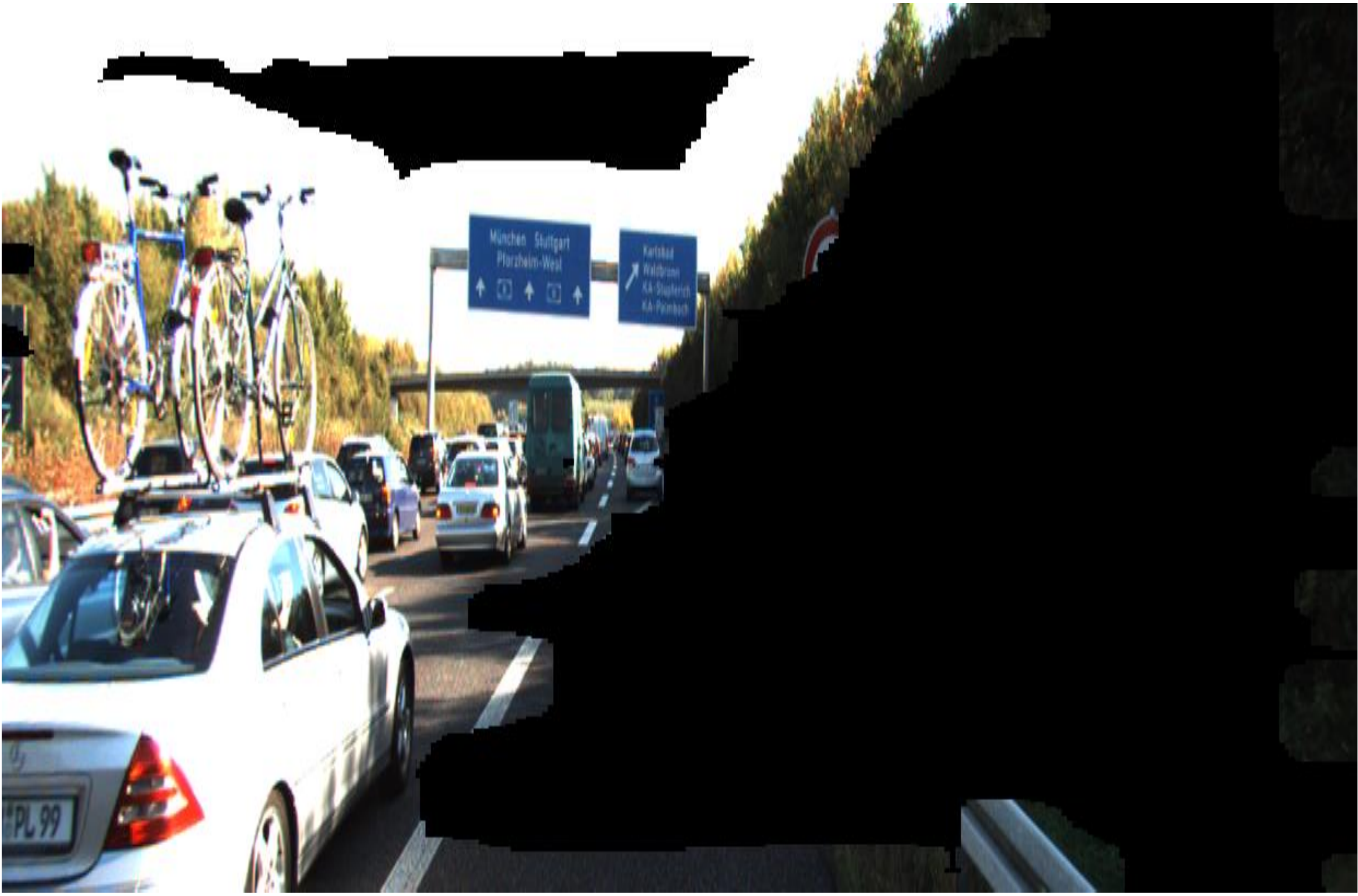}}
	}
\caption{Input and its intermediate and resultant images of the SRVS model: a) input image;  b) saliency map; c) object mask; and (d) masked result.}
\label{fig:SRVS}
\end{figure}

In the proposed EODF, it is assumed that AVs have hardware with limited resources (e.g., a processor has low clock speed and does not support multi-thread processing). Thus, we adopt the SRVS model~\cite{HouCVPR2007} for salient region detection, which can efficiently operate on low-performance computational units.\footnote{Note that other methods with low computational complexity can be used instead of the SRVS model. However, even though we have tested lots of methods (e.g., binarized normed gradients (BING)), it is observed that the SRVS model can achieve the best performance. The representative comparison with BING will be elaborated in Table~\ref{tab:speed}.} Figure~\ref{fig:SRVS} shows the procedure of obtaining RoIs based on SRVS model. Figure~\ref{fig:SRVS}(a) represents an input image. In SRVS model, the frequency spectrum of the image is first obtained by the Fourier transform. Then, to obtain the low-frequency spectrum, the frequency spectrum is smoothed by using a local average filter.\footnote{The low-frequency spectra of any images have a similar tendency~\cite{HouCVPR2007}, which implies that the low-frequency spectrum of any general images can be approximated by just smoothing the spectrum.} Note that the low frequency in images represents that pixel values change slowly over space, which means that the contents with low frequency are probably the background regions. After obtaining the low-frequency spectrum, high-frequency regions called \textit{spectral residual} (i.e., salient regions of an image) are driven by subtracting the low-frequency spectrum from the original spectrum of the input. By using the inverse Fourier transform, we can re-transform  the spectral residual into the image domain, which results in a saliency map as shown in Fig.~\ref{fig:SRVS}(b). Then, we can obtain an object mask (see Fig.~\ref{fig:SRVS}(c)) on the basis of the saliency map and a saliency threshold. Specifically, if the regions in the saliency map have larger (smaller) values than the saliency threshold, the corresponding regions in the object mask are set to one (zero). Note that these regions can be considered as object-expected and non-object-expected regions, respectively. Therefore, by multiplying the object mask with the input image, we can extract the RoIs of the input image, in which objects such as vehicles and pedestrians are well included, as shown in Fig.~\ref{fig:SRVS}(d).

\subsection{Object Detection}

Object detection is one of the most challenging problems owing to the varying scales and poses of objects in a scene. Various attempts using low-level features, such as colors and intensity gradients, have been made to overcome this problem. Recently, with a decade of advances in convolutional neural networks~(CNNs), vital clues for the solution of the problem have been provided.\footnote{Even emotion cognition and disease diagnosis can be realized with recent CNNs~\cite{Chen20_LabelLess,chen2017deep}.}

CNN-based object detection can be divided into two categories: 1) two-stage detectors and 2) one-stage detectors. The initial two-stage detectors recognize the image regions of the object candidates provided from a region proposal module using deep CNN classifiers. In the recent two-stage detectors, the region proposal module is incorporated into an object detection system as an end-to-end CNN model. On the other hand, one-stage detectors predict the image regions of the object candidates directly without a region proposal module. This allows time-efficient processing for real-time applications. The single-shot multi-box detector~(SSD)~\cite{LiuECCV2016} is a representative example of one-stage detectors. SSD utilizes a feature pyramid, whose level consists of CNN layers responsible for detecting objects within a certain size range. By adopting the pyramid designs, the SSD provides detection performance comparable to that of Faster R-CNN with low complexity (i.e., low latency).

In this article, we employ the receptive field block~(RFB)~network~(RFBNet)~\cite{LiuECCV2018} (that is a representative one-stage detector) in edge clouds for object detection, because it achieves state-of-the-art performance on challenging object detection benchmarks with real-time speed and has a robust deep network architecture operating well even when some parts of objects are removed from an input image.\footnote{Note that, in the EODF, when the channel condition is poor, object detection should be conducted with the compressed image having a possibility that some parts of objects are removed.} In RFBNet, the network architecture of the SSD is extended by replacing the top convolution layers of SSD with a specific module, namely the RFB module. In the RFB module, multiple branches with different receptive field sizes are combined. Inspired by the behavior of receptive fields in human visual systems, object features at different scales are extracted in each branch of the RFB module and aggregated to produce enriched object information.%\footnote{Detailed network architectures can be found in~\cite{LiuECCV2018}.}

\section{Edge Network-Assisted Object Detection Framework}
\label{Sec:EODF}

\begin{figure*}
\includegraphics[width=\linewidth]{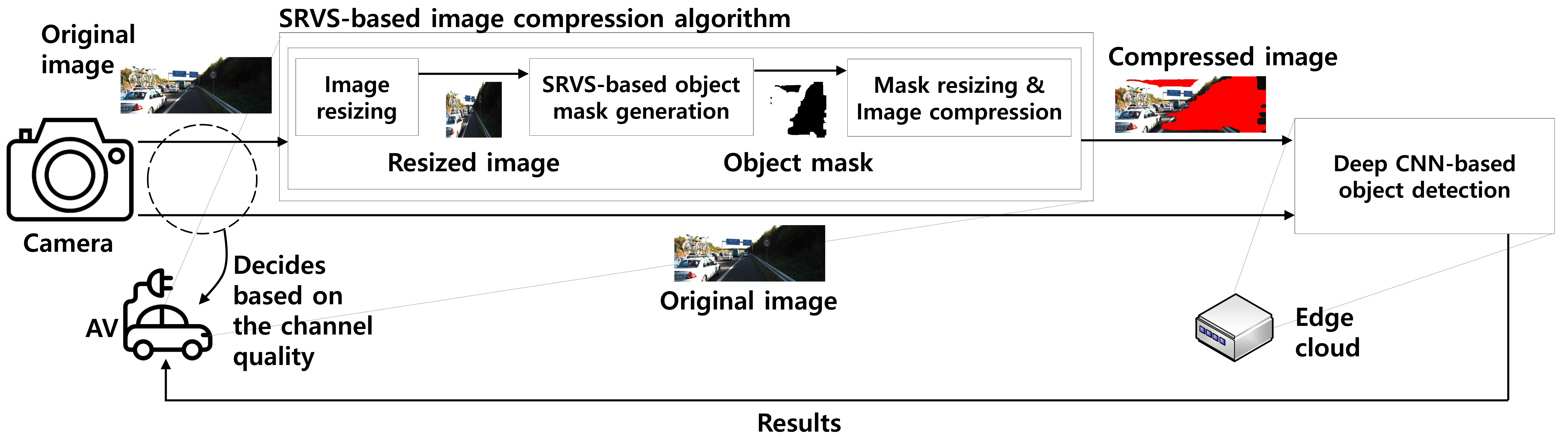} \centering
\vspace{-1.5cm}
\caption{Implementation overview. For visualization, the discarded regions in the compressed image are colored in red.}
\label{fig:overview}
\end{figure*}

%\footnote{AV can check the channel quality based on a channel quality indicator (CQI) index included in the CSI message.}

In this section, we elaborate on the proposed EODF for autonomous driving. Figure~\ref{fig:overview} shows the implementation overview of the EODF. AVs check the channel quality by exchanging control messages regarding the channel status information (CSI) with a base station (BS). Based on the channel quality, the AVs determine whether to compress the image or not. If the channel quality is sufficiently good, i.e., the AVs expect to receive the object detection result of the current image within the inter-frame duration, the AVs do not compress the image. On the other hand, if the channel quality is poor, i.e., the AVs estimate that they cannot receive the result of the current image within the inter-frame duration, the AVs compress the image using an SRVS-based image compression algorithm. In the SRVS-based image compression algorithm, a saliency map of an input image is first generated. To boost the speed of the generation, we resize the input image of the SRVS model with a smaller size (e.g., ${64}\times{64}$). After that, a binary object mask, where the zeros and ones represent the non-object-expected and object-expected regions, respectively, is created by binarizing the saliency map with a predefined saliency threshold. Specifically, if a pixel value of the saliency map is higher than the saliency threshold, the corresponding point of the binary object mask is set to one. Otherwise, its corresponding point is set to zero.\footnote{If a higher saliency threshold is used, the object mask covers smaller areas where objects are likely to be included. Therefore, the image compression rate can increase with a higher saliency threshold.} After making the binary object mask, it is up-sampled into the size of the original image. Then, by multiplying the input image with the up-sampled binary object mask, the data size of the input image is reduced (i.e., the input image is compressed), because the non-object-expected regions of the input image become zeros after the multiplication. Then, the AVs transmit either the compressed image or the original one to the edge cloud. After receiving the image, irrespective of whether it is compressed or not, the edge cloud conducts object detection by means of RFBNet and transmits the results (i.e., bounding box coordinates of objects and their categories) to the AVs. Note that, even though the edge cloud receives the compressed image, the edge cloud conducts object detection without any decompression procedure (i.e., the compressed image is used as an input of RFBNet). However, since the important contextual information might be contained in salient regions and the image is compressed without huge damage on salient regions, the accuracy of object detection does not significantly decrease.

\begin{figure}
\includegraphics[width=12cm]{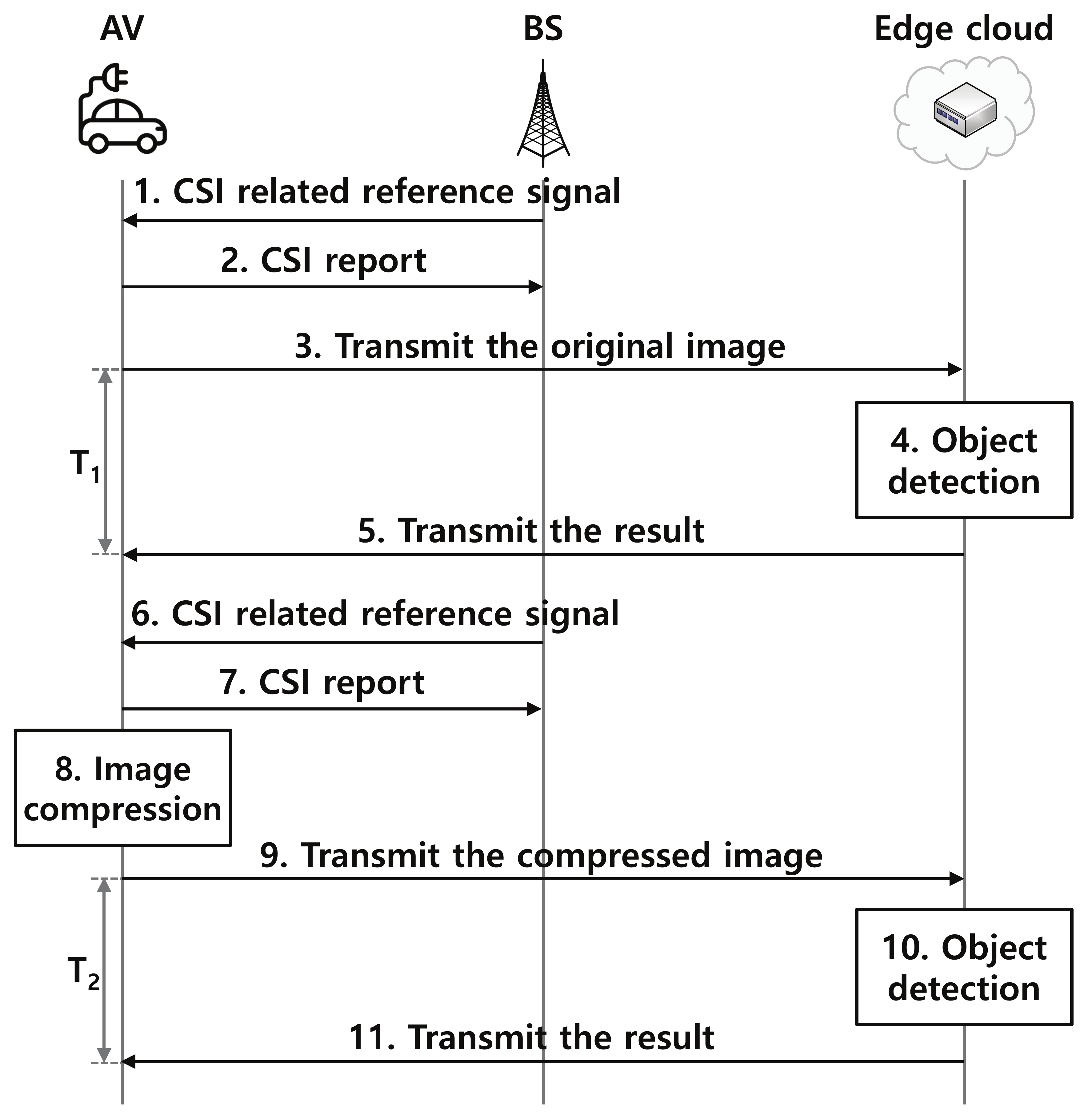} \centering
\vspace{-0.4cm}
\caption{Operation example of the EODF.}
\label{fig:Operational_example_EODF}
\end{figure}

Figure~\ref{fig:Operational_example_EODF} shows an operational example of the EODF. After receiving the CSI related reference signal from the BS~(step 1), the AV checks the channel quality and transmits the CSI report to the BS~(step 2). If it is assumed that the channel quality is sufficient to receive the result within the inter-frame duration, the AV transmits the original image to the edge cloud through the BS~(step 3). After performing object detection, the edge cloud transmits the object detection results to the AV~(steps 4-5). For the next CSI related reference signal~(step 6), the AV checks the channel quality and transmits the CSI report to the BS again~(step 7). In this case, since the AV determines that the channel quality is not enough good to receive the result within the inter-frame duration when transmitting the original image, the AV compresses the image based on the SRVS model~(step 8). Then, the AV can transmit the compressed one with a lower latency due to the smaller size of image data~(step 9). After receiving the compressed image, the edge cloud performs object detection and transmits the result to the AV~(steps 10-11).

\section{Evaluation Results}
\label{Sec:Evaluation}

In this section, we evaluate the performance of our preliminary implementation. For performance evaluation, we compared the proposed framework, EODF, with a conventional edge-assisted object detection framework (denoted by CONV) where the AV always transmits the original image to an edge cloud without any consideration of the channel condition. Meanwhile, since the objective of the EODF is to guarantee a lower latency than the inter-frame duration while maintaining average precision above a sufficient level, the average outage probability, representing the average probability that results are not received within the inter-frame duration, and average precision are used as the performance measures of the EODF. We used the KITTI dataset~\cite{GeigerCVPR2012}, which is a challenging real-world autonomous driving dataset. Additionally, we implemented the SRVS-based image compression algorithm in low-power embedded systems (i.e., Raspberry Pi 3 B+ and Intel i5-8250U CPU) and deep CNN-based object detection algorithm (i.e., RFBNet) in the edge cloud (i.e., V100 GPU). Meanwhile, we assumed 5G cellular systems. Specifically, the number of aggregated component carriers, the maximum number of MIMO layers, and the scaling factor were set to 2, 4, and 1, respectively. In addition, the bandwidth was assumed to be $20$ MHz and the channel quality changed randomly. Meanwhile, the threshold for the channel quality was set to $7$ CQI index.

\subsection{Compression and Object Detection Speeds}

\begin{table}
    \caption{Compression speed~(fps) comparison between BING~\cite{Cheng14} and SRVS model~\cite{HouCVPR2007} and object detection speed~(fps) of RFBNet~\cite{LiuECCV2018} for different input sizes}
	\begin{center}
		\scalebox{1.0}
        {
            \begin{tabular}{|l|c c|c c|c|}
                \hline
                \multirow{2}{*}{Input size} & \multicolumn{2}{c|}{Raspberry Pi 3 B+} & \multicolumn{2}{c|}{Intel i5-8250U CPU} & V100 GPU \\
                \cline{2-6}
                 & BING & \multicolumn{1}{c|}{SRVS} & BING & \multicolumn{1}{c|}{SRVS} & RFBNet \\
                \hline\hline
                ${512}\times{512}$ & 0.2 & 45 & 2 & 233 & 59 \\
                \hline
                ${256}\times{256}$ & 0.5 & 123 & 6 & 748 & 91 \\
                \hline
                ${128}\times{128}$ & 0.7 & 172 & 18 & 1,497 & 125 \\
                \hline
                ${64}\times{64}$ & 1.1 & 240 & 51 & 2,862 & 200 \\
                \hline
            \end{tabular}
        }
    \end{center}
    \label{tab:speed}
\end{table}

Since the compression speed in low-power embedded systems and object detection speed in the edge cloud influence the average outage probability, we measured these speeds. Table~\ref{tab:speed} presents a summary of the compression speed in low-power embedded systems and object detection speed in the edge cloud according to the input sizes. As listed in Table~\ref{tab:speed}, the compression speed in low-power embedded systems is understandably boosted as the input size becomes smaller. As suggested in~\cite{HouCVPR2007}, we also adopted an input size of ${64}\times{64}$, which can be operated at 240 fps and 2,862 fps on Raspberry Pi 3 B+ and a single-core CPU, respectively. It is worthwhile to note that the use of representative RoI extraction methods such as BING~\cite{Cheng14} can be considered. In BING, a convolution operator is learned using the norm of the image gradients, and then decomposed into a few atomic operations such as addition, bit-wise shift, etc., to boost the operation speed. However, in our scenario, we assumed that AVs have very limited computational resources; therefore, even BING is computationally demanding. For example, as shown in Table~\ref{tab:speed}, BING operates at 1.1 fps, 0.7 fps, 0.5 fps, and 0.2 fps for the input sizes of $64 \times 64$, $128 \times 128$, $256 \times 256$, and $512 \times 512$, respectively. Meanwhile, object detection speed in the edge cloud is sufficiently small for the real-time service irrespective of the input size, as shown in the last column of Table~\ref{tab:speed}.

Table~\ref{tab:speed} shows that AVs in the proposed EODF can successfully operate on low-performance computing units (i.e., Raspberry Pi 3 B+ and Intel i5-8250U CPU). Without the EODF, AVs should have expensive computational units (e.g., NVIDIA Tesla V100 GPU, whose cost is about 200 times more expensive than Raspberry Pi B+) to conduct the object detection in real-time. In this context, we can expect that the price of AVs is significantly reduced when using the EODF.

\subsection{Average Outage Probability}

%One of the prerequisite conditions of successful real-time OD based on edge clouds is that the overall latency for OD is smaller than the IFD.

\begin{figure}
\includegraphics[width=9cm]{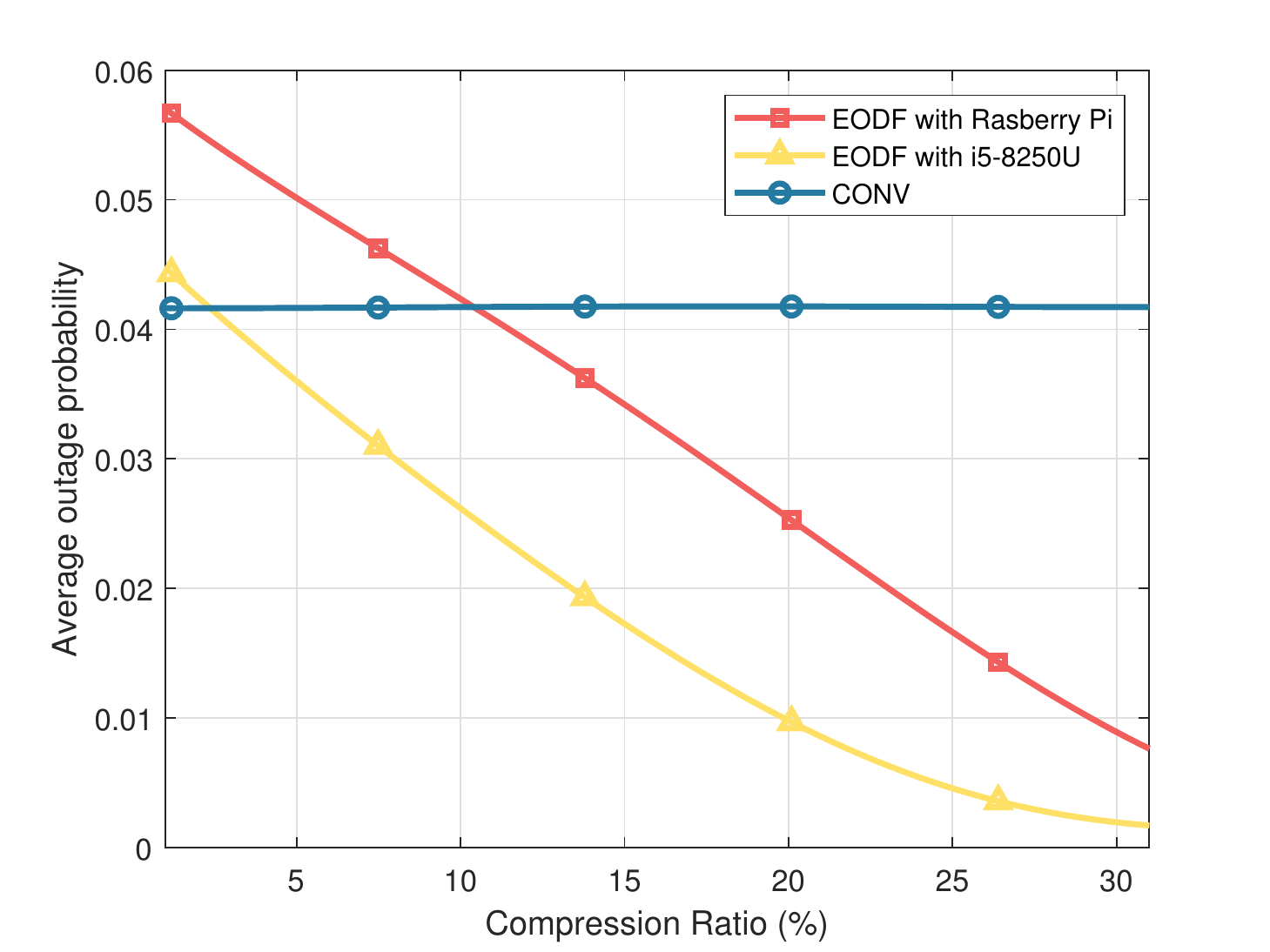} \centering
\vspace{-0.5cm}
\caption{Effect of compression ratio on average outage probability.}
\label{fig:Average_Outage_Prob}
\vspace{-0.2cm}
\end{figure}

%\footnote{Note that the overall latency for OD in the EODF includes compression latency in low-power embedded systems, transmission latency of compressed image to the edge cloud, OD detection latency in the edge cloud, and transmission latency of results to AVs.}

Based on these real measurement values, we conducted extensive simulations to evaluate the average outage probability. The average outage probability is calculated with a criterion of 30 fps. Figure~\ref{fig:Average_Outage_Prob} shows the average outage probabilities of EODF and CONV according to the compression ratio. From Fig.~\ref{fig:Average_Outage_Prob}, it can be shown that the average outage probability of EODF decreases with the increase of the compression ratio. Especially when the compression ratio is higher than 10\%, the EODF even with Raspberry Pi 3 B+ can achieve better performance than CONV. This is because a higher compression ratio means that the AVs in the EODF can transmit a smaller sized image to the edge cloud. In this situation, the compressed image can be delivered to the edge cloud with low latency even though the channel quality is relatively poor. Note that the effect of the compression latency can be more dominant than that of the reduced transmission latency, especially when the EODF operates on Raspberry Pi 3 B+ and the compression ratio is small (i.e., smaller than 10\% in our simulation settings). In this situation, the average outage probability of EODF with Raspberry Pi 3 B+ can be higher than that of CONV.

\begin{figure}[]
\centering
	{
	\subfigcapskip = -15pt
		\subfigure[]{
		\includegraphics[width =0.48\textwidth]{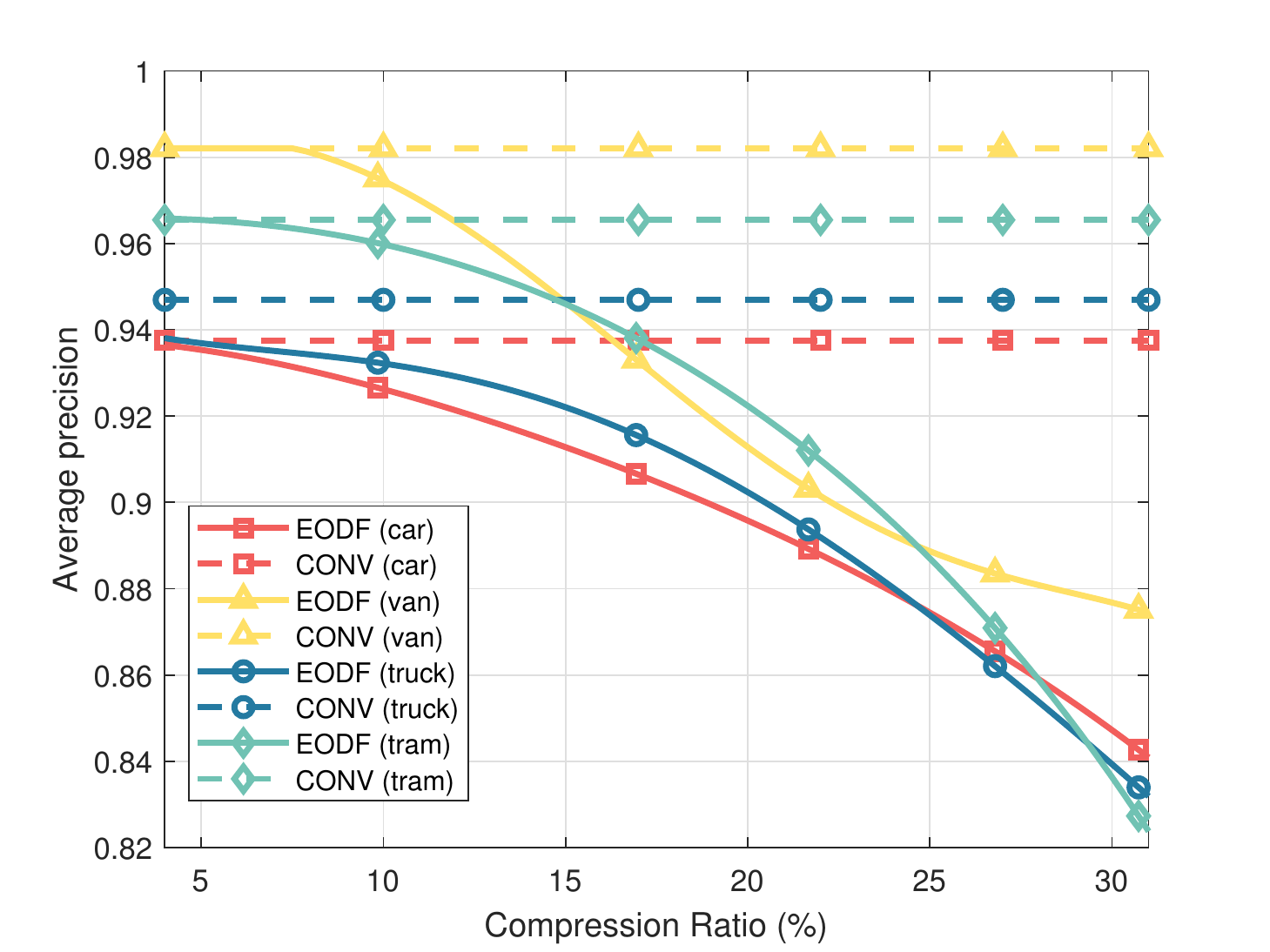}}
		\subfigure[]{
		\includegraphics[width=0.48\textwidth]{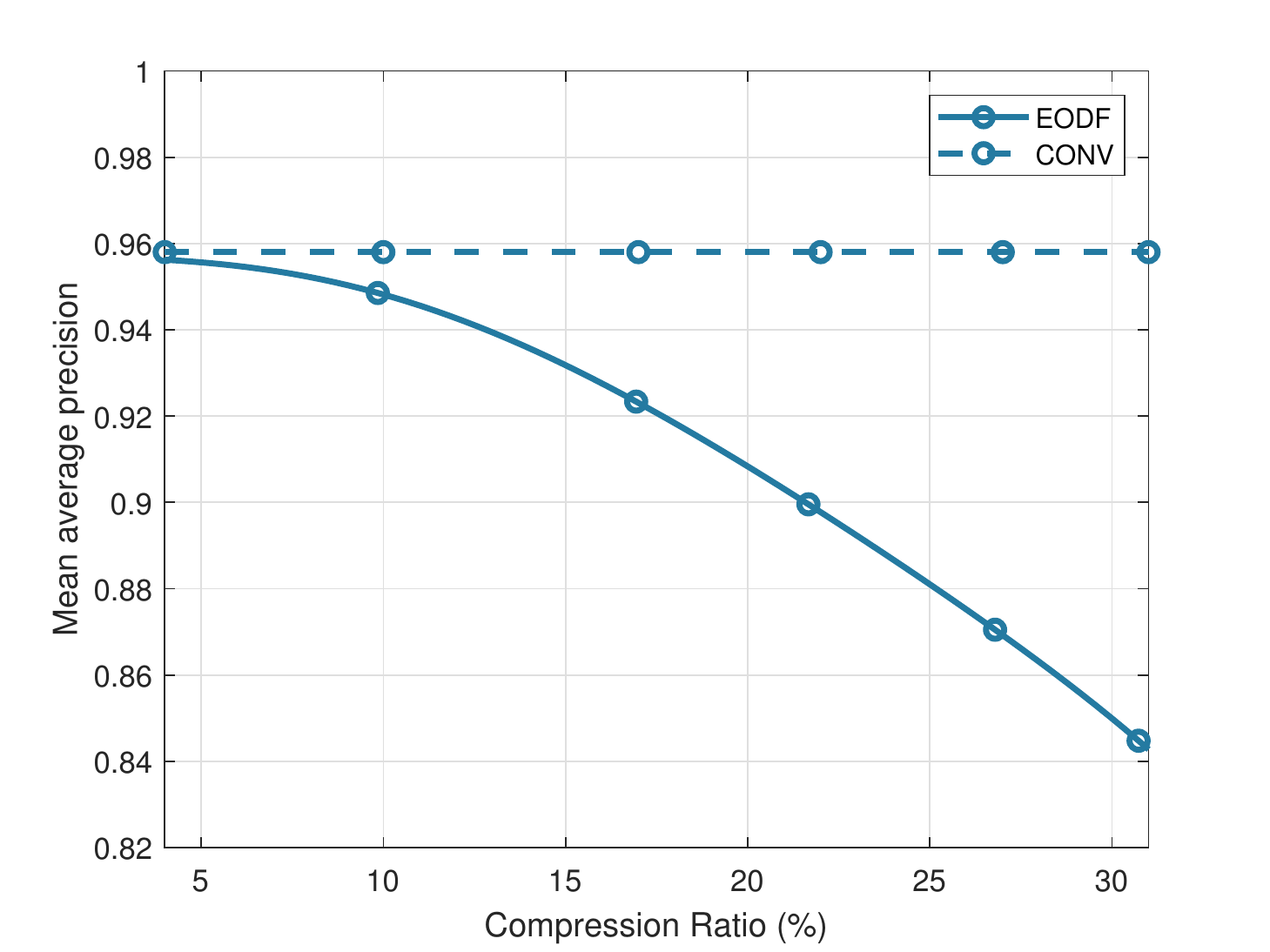}}\\
	}
    \caption{Effect of compression ratio: a) Average precision; b) Mean average precision.}
    \label{fig:AP}
\end{figure}

%\begin{figure}
%\begin{minipage}{0.5\textwidth}
%    \includegraphics[width=\linewidth]{Figure/each_ap.eps}
%    \centering (a)
%\end{minipage}
%\hfill
%\begin{minipage}{0.5\textwidth}
%    \includegraphics[width=\linewidth]{Figure/mean_ap.eps}
%    \centering (b)
%\end{minipage}
%\caption{Effect of compression ratio: a) Average precision; b) Mean average precision.}
%\label{fig:AP}
%\end{figure}

\subsection{Average Precision}
\label{ssec:mAP}

Figure~\ref{fig:AP}(a) shows the compression ratio versus~(vs) accuracy (i.e., average precision) trade-off for each object category (i.e., car, van, truck, and tram). Similarly, Figure~\ref{fig:AP}(b) depicts the compression ratio vs accuracy (i.e., mean average precision) trade-off for the overall objects. From Fig.~\ref{fig:AP}(a), it can be observed that with a relatively small compression ratio~(under 10\%), the APs of EODF are comparable with those of CONV. Specifically, when the compression ratio is around 7.5\%, APs of EODF and CONV for van and tram are the same. This can be explained as follows. When a small compression ratio is used, SRVS model can remove the only low-textured regions such as the sky and road. Thus, with the small compression ratio~($<7.5\%$), any information for object detection may not be lost and the performance (i.e., average precision) of object detection is maintained.

Meanwhile, from Figs.~\ref{fig:AP}(a) and (b), it can be seen that the average precision of each object and mean average precision of all objects decrease as the compression ratio increases. Even though the SRVS model attempts to extract the accurate RoIs in which the objects are contained, as the number of pixels to be removed increases~(i.e., a higher compression ratio is exploited), the risk of missing pixels corresponding to objects increases, especially for the object regions whose color or textures are similar to the background. In this situation, the lost contextual information on objects can affect the performance of identifying an object. However, even with the high compression ratio of 32\%, the EODF still achieves an average precision of over 82\%  for all categories, including car, van, truck, and tram, as shown in Fig.~\ref{fig:AP}(a), and the overall performance is a mean average precision of over 84\%, as shown in Fig.~\ref{fig:AP}(b). We believe that this is an acceptable accuracy for real-world autonomous driving applications.

From Figs.~\ref{fig:Average_Outage_Prob} and~\ref{fig:AP}, it can be observed that an optimal compression ratio exists. Smaller average outage probability can be obtained with a higher compression ratio (see Fig.~\ref{fig:Average_Outage_Prob}). However, a significantly higher compression ratio can cause information loss, which results in performance degradation of object detection (i.e., degradation of average precision) (see Fig.~\ref{fig:AP}). To conclude, the level of average precision that should be maintained for optimal AV operations needs to be evaluated. Then, the compression ratio can be set to the maximum value to guarantee the evaluated average precision level.

\section{Future Research Directions}
\label{Sec:FRD}

\subsection{Research on Performance Improvement of RoI Extraction Methods}

Although the object detection performance with image compression (over mean average precision of 84\% with a compression ratio of 30\%) is reasonably acceptable if the objects and backgrounds have similar textures, a part of object regions could be removed in the image compression step. Thus, a more accurate salient region detection method can be developed to improve the performance of the EODF. One of the possible methods is updating the smoothing filter of the SRVS-based image compression algorithm. Specifically, when the channel quality is good, the edge cloud can receive the full image data from the AV. Then, the edge cloud can compress the original image via the same image compression algorithm as the AVs. Because the edge cloud has the bounding box results of the original image, it can be checked whether the object regions detected in the original image disappear in the compressed image. Based on this observation, we can update the smoothing filters of the SRVS model. The edge cloud then provides AVs an updated filter to enhance the RoI extraction method.

\subsection{Research on Performance Improvement of Object Detection Methods}

In the EODF, there is a possibility of losing important contextual information of objects when the image is compressed, which can degrade the accuracy of object detection. To mitigate this problem and supplement information for object detection, the information included in previous images can be exploited. Note that, in the EODF, the edge cloud receives a series of images that may contain the same object over time. Especially when previous images are not compressed, those images have extensive information compared to the compressed one. Therefore, to improve the performance (i.e., accuracy) of object detection, a method of how to exploit such temporal information can be devised. Even though existing video object detection methods~\cite{zhu2017flow,fan2019shifting} provide some insight of using temporal information, since these methods do not consider compressed images, they cannot be directly applied to the EODF.

\subsection{Research on Security Protocol}

In edge cloud-assisted autonomous driving systems, there are two main security issues: 1) denial-of-service (DoS) attacks and 2) hijacking images. In a type of DoS attack, an attacker sends several fake images to a target edge cloud, thereby overloading it with high computational requirements. As these images are processed constantly in the target edge cloud, it becomes overwhelmed, which causes a DoS condition to legitimate AVs. Meanwhile, in the case of hijacking images, the adversary can hijack an image transmitted from AVs and place fake traffic lights, traffic signs, and traffic objects (e.g., cars or pedestrians). Then, it transmits the modified image to an edge cloud, which generates wrong results. If these wrong results are delivered to the AVs, significant disasters (e.g., traffic accidents) can occur. To prevent these issues, a security protocol should be devised. Moreover, the devised security protocol should have a short operational latency to prevent it from affecting the performance of edge cloud-assisted autonomous driving systems.

\subsection{Research on Scalability}

Even though edge clouds have high computing power, they can be overloaded when several AVs offload their tasks simultaneously to a single edge cloud. If edge clouds are overloaded, the results may not be returned within the inter-frame duration, which reduces the quality of service of users. When lots of edge clouds are deployed, this problem can be naturally resolved. However, it can increase the capital expenditure of an offloading service provider. Therefore, it is important to decide the appropriate number and locations of edge clouds to be deployed. The optimal number and locations of edge clouds can be determined by formulating an integer linear programming problem. In addition, the collaboration between these deployed edge clouds (e.g., distributed computing algorithms) is needed to appropriately distribute the offloaded tasks.

\section{Conclusion}
\label{Sec:Conclusion}

In this article, we developed an EODF for autonomous driving. In the proposed framework, AVs operate adaptively according to the channel quality. Specifically, if the channel quality is sufficient to offload an original image and receive the result from the edge cloud within the frame rate, the AVs transmit the original image to the edge cloud. Otherwise, the AVs compress the image data by using the SRVS-based image compression algorithm and transmit it. In so doing, the transmission latency can be significantly reduced, and therefore the AVs probably receive the result of object detection within the inter-frame duration. Evaluation results demonstrated that our framework can achieve real-time object detection with satisfactory accuracy. In our future works, we will devise a method to decide the optimal threshold of the channel quality. Moreover, we will conduct additional comparative evaluations on multi-AV environments with various types of tasks.

\section*{Acknowledgment}

This research was supported by National Research Foundation~(NRF) of Korea Grant funded by the Korean Government~(MSIP)~(No. 2019R1C1C1004352).

\bibliographystyle{IEEEtran}
\bibliography{IEEEabrv,CODF}

Seung-Wook Kim (swkim@dali.korea.ac.kr) received B.S. and Ph.D. from the School of Electrical Engineering, Korea University, Seoul, Korea, in 2012 and 2019, respectively. He is currently a staff researcher in Samsung Advanced Institute of Technology, Suwon, Korea. From 2018 to 2019, he was a research professor in Semiconductor Research Center, Korea University, Seoul, Korea. His research interests include computer vision, image processing, machine learning, and deep learning.

\vfill

Keunsoo Ko (ksko@mcl.korea.ac.kr) received the B.S. degree in electrical engineering from Korea University, Seoul, South Korea, in 2017, where he is currently pursuing the Ph.D. degree. His current research interests include computer vision and machine learning.

\vfill
\newpage

Haneul Ko (heko@korea.ac.kr) [M'19] received B.S. and Ph.D. from the School of Electrical Engineering, Korea University, Seoul, Korea, in 2011 and 2016, respectively. He is currently an assistant professor in the Department of Computer Convergence Software, Korea University, Sejong, Korea. From 2017 to 2018, he was with the Smart Quantum Communication Research Center, Korea University, Seoul, Korea, and a visiting Postdoctoral Fellow at the University of British Columbia, Vancouver, BC, Canada. From 2016 to 2017, he was a Postdoctoral Fellow in mobile network and communications, Korea University, Seoul, Korea. His research interests include 5G networks, network automation, mobile cloud computing, SDN/NFV, and Future Internet.

\vfill

Victor C. M. Leung (vleung@ieee.org) [S'75, M'89, SM'97, F'03] is a distinguished professor of computer science and software engineering at Shenzhen University. He is also an emeritus professor of electrical and computer engineering at the University of British Columbia (UBC), where he held the positions of professor and the TELUS Mobility Research Chair until the end of 2018. His research is in the broad areas of wireless networks and mobile systems. He has co-authored more than 1200 journal/conference papers and book chapters. He is serving on the editorial boards of IEEE Transactions on Green Communications and Networking, IEEE Transactions on Cloud Computing, IEEE Access, IEEE Network, and several other journals. He received the IEEE Vancouver Section Centennial Award, 2011 UBC Killam Research Prize, 2017 Canadian Award for Telecommunications Research, and 2018 IEEE TGCC Distinguished Technical Achievement Recognition Award. He co-authored papers that won the 2017 IEEE ComSoc Fred W. Ellersick Prize, 2017 IEEE Systems Journal Best Paper Award, and 2018 IEEE CSIM Best Journal Paper Award. He is a Fellow of IEEE, the Royal Society of Canada, Canadian Academy of Engineering, and Engineering Institute of Canada.

\vfill

\end{document}